\documentclass[a4paper,12pt]{article}
\usepackage{graphicx,epsf}
\usepackage{amsbsy}
\usepackage{color}
\title{Perturbations in generalized multi-field inflation}
\author{David Langlois$^{1,2}$\footnote{langlois@apc.univ-paris7.fr} \,  and S\'ebastien Renaux-Petel$^{1}$\footnote{renaux@apc.univ-paris7.fr}   \\
{\small {}}\\
{\small ${}^1${\it APC (Astroparticules et Cosmologie),}}\\
{\small {\it
UMR 7164 (CNRS, Universit\'e Paris 7, CEA, Observatoire de
Paris)}}\\
{\small {\it  10, rue Alice Domon et L\'eonie Duquet,
 75205 Paris Cedex 13, France}}\\
 {\small ${}^2${\it Institut d'Astrophysique de Paris (IAP),}}\\
{\small {\it 98bis Boulevard Arago, 75014 Paris, France;  }}\\}

\begin{document}

\date{\today}
\maketitle

\def\half{\frac{1}{2}}
\def\beq{\begin{equation}}
\def\eeq{\end{equation}}
\newcommand{\bea}{\begin{eqnarray}}
\newcommand{\eea}{\end{eqnarray}}
\def\Tdot#1{{{#1}^{\hbox{.}}}}
\def\Tddot#1{{{#1}^{\hbox{..}}}}
\def\p{\phi}
\newcommand{\dn}[2]{{\mathrm{d}^{{#1}}{{#2}}}}
\def\PX{P_{,X}}
\def\s{\sigma}

\begin{abstract}
We study the linear perturbations of multi-field inflationary models governed by a Lagrangian which is a general function of the scalar fields and of a global kinetic term combining their spacetime gradients with an arbitrary field space metric. Our analysis includes \textit{k}-inflation, Dirac-Born-Infeld inflation and its multi-field extensions which have been recently studied. For  this general class of models, we calculate the action to second order in the linear perturbations. We decompose the perturbations into an adiabatic mode, parallel to the background trajectory, and entropy modes. We show that all the entropy modes propagate with the speed of light whereas the adiabatic mode propagates with an effective speed of sound. We also identify the specific combination of entropy modes which sources the curvature perturbation on large scales. We then study in some detail the case of two scalar fields: we write explicitly the equations 
of motion for the adiabatic and entropy modes in a compact form and discuss their quantum fluctuations and primordial power spectra.

\end{abstract}

\newpage

\section{Introduction}

Inflation has now become a standard paradigm for describing the physics of the very early universe, but the nature of the field(s) responsible for inflation remains an open question. The hope is that future cosmological observations, in particular those of the CMB, will be able to rule out large classes of models and give some hints on the underlying physics. 

In the last few years, intensive effort has been devoted to trying to connect string theory and inflation (for recent reviews, see e.g. \cite{McAllister:2007bg}-\cite{HenryTye:2006uv}). For simplicity,  most studies of string inflation have considered a single effective scalar field. 
However, in the low-energy limit of string theory,  many  scalar fields are present and several of them 
could thus  play a dynamical role during inflation. This would affect the generation of primordial perturbations.   
For instance, whereas in single-field inflation, the curvature perturbation is conserved on large scales\footnote{see however \cite{Leach:2001zf} and \cite{Jain} for exceptions.},  
 the curvature perturbation in multi-field inflation can generically be modified on large scales, because it is sourced by entropy (or
 isocurvature) perturbations.  This feature was first pointed out in \cite{sy} in the context of Jordan-Brans-Dicke type gravity where the gravitational sector contains a scalar field. This effect has also been illustrated 
 recently \cite{Lalak:2007vi,Brandenberger:2007ca} in the context of specific inflationary models based on string theory constructions, and shows that  the
restriction to an effectively single-field scenario, despite its appealing simplicity, might be misleading as the final curvature perturbation, which will be eventually observed, can sometimes originate mainly from the entropy modes.
 
In multi-field inflation, there is also the possibility, depending on the reheating scenario, to produce, after inflation, both adiabatic and isocurvature perturbations, which can be correlated, as first pointed out in \cite{Langlois:1999dw}.  The CMB measurements have shown that the primordial perturbations are mainly adiabatic but a small amount of isocurvature modes is still allowed  by the data (see e.g. \cite{Bean:2006qz}-\cite{Komatsu:2008hk} for the most recent analyses). 
 
In this context, a lot of works have been devoted to multi-field inflation. Although many of these studies usually assume standard kinetic terms for all the scalar fields involved,   several   consider  multi-field inflation  with non-standard kinetic terms described by a non-trivial metric $G_{IJ}(\phi^K)$ in field space, so that the kinetic part of the Lagrangian 
is of the form 
$${\cal L}_{\rm kin}=-\frac{1}{2} G_{IJ}\partial_\mu\phi^I\partial^\mu\phi^J.
$$ 
More sophisticated types of non-canonical kinetic terms can also be envisaged. This is in particular the case for an interesting scenario based on string theory, in which the inflaton is a scalar field characterizing the position of a probe brane moving in a warped background geometry.  This scenario has been named DBI (Dirac-Born-Infeld) inflation \cite{st03,ast04}, as the inflaton field is governed by a Dirac-Born-Infeld action. The latter can be seen as a particular case of the more general framework of  \textit{k}-inflation \cite{ArmendarizPicon:1999rj,Garriga:1999vw}, where the action is an arbitrary function of the inflaton and of the square of its spacetime gradient.
Two recent works \cite{Easson:2007dh,Huang:2007hh} have studied the perturbations in effectively multi-field DBI inflation, where the extra fields correspond  to the angular degrees of freedom of the moving brane  \cite{Kehagias:1999vr}-\cite{Easson:2007fz}.

The purpose of the present work is to analyse a very large class of multi-field models,  which can be described by an
action of the form
\beq
S =  \int {\rm d}^4 x \sqrt{-g}\left[\frac{R}{16\pi G}   +   P(X,\phi^I)\right] 
\eeq
where $P$ is an arbitrary function of $N$ scalar fields and of the  kinetic term
\beq
X=-\half G_{IJ} \nabla_\mu \p^I  \nabla^\mu \p^J.
\eeq
where $G_{IJ} \equiv G_{IJ}(\p^K)$ is an arbitrary metric on the $N$-dimensional field space.
 This can be seen as a generalization of the Lagrangian  of \textit{k}-inflation \cite{ArmendarizPicon:1999rj} to
 the case of  several scalar fields. This also includes DBI inflation and 
 its  multi-field extensions studied in \cite{Easson:2007dh,Huang:2007hh}. 

In this work, we expand  the above action up  to second order in the linear perturbations.
 The  second order action can be used to derive the classical equations of motion for the perturbations. It is also the starting point to calculate the spectra of the primordial perturbations generated from the vacuum quantum fluctuations of the scalar fields during inflation.  
  
  In order to analyse the equations of motion, we decompose the linear perturbations into the (instantaneous) adiabatic perturbation, i.e. the perturbation parallel to the background trajectory in field space, and the (instantaneous) entropy perturbations, which are orthogonal (with respect to the field space metric) to the trajectory. 
  We find, quite generically, that the equation of motion for the adiabatic perturbation is a wave equation which depends on an effective sound speed $c_s$ while all the entropy perturbations obey a wave equation involving the speed of light. This shows that the property that adiabatic and isocurvature perturbations propagate with different speeds, pointed out in \cite{Easson:2007dh} for the particular case of two-field DBI inflation, is valid for the large class of multi-field models studied here,  whatever the specific dependence of the action on the kinetic term, and whatever the number of scalar fields involved. For more than two scalar fields, the entropy sector contains several degrees of freedom but they all satisfy a light-like wave equation.
  
We also analyse the evolution of the curvature perturbation on large scales, and show that it is sourced by a specific combination of entropy perturbations, which can be decomposed into a term corresponding to the bending of the background trajectory in field space, as in multi-field inflation with canonical kinetic terms, and an additional term, which is present only for non-trivial functions $P$. 
 We finally specialize our analysis to the case of two fields and  compare our results with previous works. 

The plan of this paper is the following. In the next section,  the background equations of motion are given.  Section 3 is devoted to the derivation of the second-order action for the linear perturbations. In the subsequent section, we analyse the (comoving) curvature perturbation and its  large-scale evolution. In Section 5, we  restrict our analysis  to two scalar fields and study in detail the equations of motion for the adiabatic and isocurvature components. Using the second-order action, we then discuss, in 
 Section 6, the quantization of the perturbations in the simple case where the adiabatic and isocurvature modes are decoupled. We conclude in the final section and 
present in an appendix the application of our formalism  to the DBI case. 

\section{Background}
As discussed in the introduction, our starting point is  the action
\beq
S =  \int {\rm d}^4 x \sqrt{-g}\left(\frac{1}{2} R  +  P(X,\phi^I)\right) \label{action}
\eeq
with 
\beq
X=-\half G_{IJ} \nabla_\mu \p^I \nabla^\mu \p^J\,.
\eeq
where we have set  $8 \pi G=1$ for simplicity. Throughout this paper, we will use the implicit summation rule on the field indices $I, J, \dots$.

The energy-momentum tensor, derived from (\ref{action}),   is of the form
\beq
T^{\mu \nu}=P\,  g^{\mu \nu}+\PX \, G_{IJ}\, \nabla^{\mu} \p^I \nabla^{\nu} \p^J\,,
\label{Tmunu}
\eeq
where $\PX$ is the partial derivative of $P$ with respect to $X$.
The equations of motion for the scalar fields, which can be seen as generalized Klein-Gordon equations, are obtained from the variation of the action with respect to $\phi^I$. One finds
\beq
\nabla_{\mu}(\PX G_{IJ} \nabla^{\mu} \p^J)-\half \PX (\nabla_\mu \phi^K)(\nabla^\mu \phi^L)
\partial_I G_{KL}+P_{,I}=0 \,.
\label{KG-general}
\eeq
where $P_{,I}$ denotes the partial derivative of $P$ with respect to $\phi^I$.

In a spatially flat FLRW (Friedmann-Lema\^itre-Robertson-Walker) spacetime, with metric
\beq
ds^2=-dt^2+a^2(t)d{\vec x}^2,
\eeq
the scalar fields are homogeneous and the energy-momentum tensor reduces to that of a perfect fluid
with energy density 
\beq
\rho=2X \PX -P\,,
\label{rho}
\eeq
and pressure $P$. 
 
The evolution of the scale factor $a(t)$ is governed by the Friedmann equations, which can be written in the form
\beq
H^2=\frac{1}{3}(2X \PX -P)\,,
\label{Friedmann1}
\eeq
and
 \beq
\dot H =- X \PX
\label{Friedmann2}.
\eeq
The equations of motion (\ref{KG-general}) for the scalar fields reduce to
\beq
\ddot \p ^I+\Gamma^I_{JK}\dot \p^J \dot \p^K +\left(3H+\frac{\dot \PX}{\PX}\right)\dot \p ^I - \frac{1}{\PX}G^{IJ} P_{,J}=0\,,
\label{KG1}
\eeq
where the $\Gamma^I_{JK}$ denote the Christoffel symbols associated with the field space metric $G_{IJ}$.
By noting that the first two terms in the above equation simply correspond to the components of the acceleration in curved coordinates 
(here in field space), which we can write as 
\beq
\mathcal{D}_t \dot \p^I \equiv \ddot \p ^I+\Gamma^I_{JK}\dot \p^J\dot\p^K\,,
\label{Dt}
\eeq
equation (\ref{KG1}) can be rewritten as 
\beq
\mathcal{D}_t \dot \p ^I +\left(3H+\frac{\dot \PX}{\PX}\right)\dot \p ^I - \frac{1}{\PX}G^{IJ} P_{,J}=0\,,
\label{KG}
\eeq
or, in an even more compact form, as
\beq
a^{-3}\mathcal{D}_t (a^3 \PX \dot{\p_I})=P_{,I}\,,
\label{KG-symetry}
\eeq
where we have used the field metric $G_{IJ}$ to lower the field index $I$, so that $\dot\p_I\equiv G_{IJ}\dot\p^J$; $\mathcal{D}_t$ acts as an ordinary time derivative on field space scalars (i.e. quantities without field space indices) and $\mathcal{D}_t G_{IJ}=0$.

The above expressions are compact but  second derivatives of the scalar fields are hidden in the term $\dot{\PX}$. It is sometimes more useful to express the equations of motion in the form
\beq
\mathcal{D}_t \dot \p ^I +\left(3Hc_s^2+\frac{c_s^2}{\PX}P_{,XK}\dot \p^K+\frac{1-c_s^2}{2 X \PX  }P_{,K} \dot \p^K\right)\dot \p ^I - \frac{1}{\PX}G^{IJ} P_{,J}=0\,,
\label{KGbis}
\eeq
 where we have introduced, as in \cite{Garriga:1999vw}, 
 \beq
 c_s^2\equiv\frac{\PX}{\rho_{,X}}=\frac{\PX}{\PX+2XP_{,XX}}\,,
\label{cs2}
\eeq
which can be interpreted as the square of an effective sound speed, as we will see later. We will also use
the dimensionless parameter 
\beq
s=\frac{\dot c_s}{H c_s}.
\eeq

\section{Dynamics of the linear perturbations}
We now study the linear perturbations about the background solution discussed in the previous section. 
\textit{A priori}, one must consider the perturbations of the scalar fields as well as the metric perturbations. The scalar metric perturbations are coupled to the  scalar field perturbations via the scalar field equations of motion and Einstein's equations. One could write down directly the linearized version of these equations. However, since we will be interested by the quantum fluctuations of the perturbations during the inflationary phase, it is useful to compute the action at  second order in the perturbations. From this action, one can easily derive the equations of motion for the linear perturbations. One can also determine the normalization of the vacuum quantum fluctuations, and therefore the amplitude of  the primordial cosmological perturbations. 

In order to quantize the system, it is useful to write the second-order action as an expression depending only
on the true physical degrees of freedom. This can be done by using the constraints to simplify the second-order action but this is a tedious procedure \cite{mfb}. A quicker method  uses the constraints, within the Hamiltonian formalism, as Hamilton-Jacobi equations in order to identify automatically the physical degrees of freedom \cite{Langlois:1994ec} (see also \cite{Anderegg:1994xq}). An even simpler procedure, introduced   in \cite{Maldacena:2002vr},  consists in writing the action in the ADM form \cite{adm} and in solving explicitly the constraints for the lapse and the shift. 

In the case of a single scalar field, two gauge choices are natural: the first is the gauge where the scalar field is spatially uniform on constant time slices; the second is the spatially flat gauge, where the spatial part of the metric is unperturbed. 
The physical perturbation is described only by the metric in the first case, and only by the scalar field perturbation in the second case. When dealing with several scalar fields, the first possibility is no longer possible and the only natural choice is to go in the spatially flat gauge, where the scalar field perturbations correspond to the physical degrees of freedom. The calculation which follows can be seen as a generalization (and a unification) 
of the second-order actions computed in  \cite{Seery:2005wm,Seery:2005gb} with the same procedure.

\subsection{ADM formalism and constraints}

In the ADM formalism \cite{Salopek:1990jq},  the metric is written in the form
\beq
ds^2=-N^2 dt^2 +h_{ij} (dx^i+N^i dt)(dx^j+N^j dt)\,
\label{metric}
\eeq
where $N$  is the lapse function and $N^i$ the shift vector.
The action (\ref{action}) then reads
\beq
S=\frac{1}{2}\int {\rm d}t {\rm d}^3 x \sqrt{h}N\,( R^{(3)}+2P)+ \half \int {\rm d}t {\rm d}^3 x \frac{\sqrt{h}}{N}(E_{ij} E^{ij}-E^2)\,,
\label{action-ADM}
\eeq
where $h=$ det$(h_{ij})$ and  $R^{(3)}$ is the Ricci curvature calculated with $h_{ij}$. The symmetric tensor $E_{ij}$, defined by 
\beq
E_{ij}=\half \dot{h}_{ij}-N_{(i|j)}
\label{Eij}
\eeq
(the symbol $|$ denotes the spatial covariant derivative associated with 
the spatial metric $h_{ij}$), is proportional to the extrinsic curvature of the spatial slices.
The function $P$ in (\ref{action-ADM}) depends on the kinetic term $X$, which can be written as 
\beq
X=\frac{1}{2N^2}G_{IJ}v^Iv^J -\frac{G_{IJ}}{2}h^{ij} \partial_i \p^I \partial_j \p^J
\label{X-ADM}
\eeq
with 
\beq
v^I=\dot \p^I -N^j \partial_j \p^I\,.
\label{speed}
\eeq
The variation of the action with respect to $N$ yields
the energy constraint,
\beq
2P-\frac{1}{N^2}(E_{ij} E^{ij}-E^2+2\PX G_{IJ}v^I v^J)=0\, ,
\label{N-constraint}
\eeq
while the variation of the action with respect to the shift $N_i$ gives 
the momentum constraint, 
\beq
  \left( \frac{1}{N} (E^j_i - E \delta^j_i) \right)_{|j} =
  \frac{\PX}{N} G_{IJ} v^I \partial_i \p^J .
  \label{Ni-constraint}
\eeq

In order to study the linear perturbations about the FLRW background, 
we now restrict ourselves to  the flat gauge, which corresponds to the choice
\beq
h_{ij}=a^2(t) \delta_{ij}.
\eeq 
The scalar fields on the corresponding flat hypersurfaces can be decomposed into 
\beq
\p^I = \p_0^I + Q^I ,
\eeq
 where the $\p_0^I$ are the spatially homogeneous background values of the fields and the $Q^I$ represent 
 the linear perturbations. In the following, we will usually drop the subscript `0' on $\p^I_0$ and simply identify $\p^I$ with the homogeneous background fields.

We also can also write the (scalarly) perturbed lapse and shift as
\beq
N = 1 + \alpha, \qquad  N_i = \beta_{,i}\, ,
\eeq
where the linear perturbations $\alpha$ and $\beta$ are determined in terms of the scalar field perturbations 
$Q^I$ by solving the linearized constraints.
At first-order,  the momentum constraint implies
\beq
  \label{alpha}
  \alpha = \frac{\PX}{2H}\dot \p_{I} Q^I\, ,
\eeq
while the energy constraint yields 
\beq
  \label{psi}
  \partial^{2}\beta = \frac{a^2}{2H} \left[ -\frac{\PX}{c_s^2}\dot{\p_I}\mathcal{D}_t Q^I -2X P_{,XI} Q^I+P_{,I} Q^I+\frac{\PX}{H}\left(\frac{X \PX}{c_s^2}-3H^2\right)\dot{\p_I} Q^I \right],
\eeq
where we have extended the notation $\mathcal{D}_t$, introduced in (\ref{Dt}),
to $Q^I$, so that 
\beq
\mathcal{D}_t Q^I \equiv \dot Q^I + \Gamma^I_{JK} \dot \p^J Q^K.
\eeq

\subsection{Second-order action}
We now expand, up to quadratic order, the action in terms of the linear perturbations. This action can be written solely in terms of the physical degrees of freedom $Q^I$ by substituting\footnote{In the ADM formalism, it is sufficient to use the  perturbed lapse and shift up to first order as their second-order parts cancel out in the action.} the expressions (\ref{alpha}) and 
(\ref{psi}) for $\alpha$ and $\beta$. In fact, it turns out that $\beta$ disappears of the second order action, after an integration by parts.  The calculation is straightforward, although somewhat lengthy. It is also convenient to regroup the terms involving derivatives of the field space metric $G_{IJ}$ into covariant expressions. 
For example, we use the identity
\begin{eqnarray}
&&\hspace*{-3.5em} \int {\rm d}t\, \dn{3}{x} \,a^3 \left[\PX G_{IJ} \dot Q^I \dot Q^J +\left( P_{,IJ}+\frac{\PX}{2}G_{KL,IJ} \dot{\p^K}\dot{\p^L} \right) Q^I Q^J
+2\PX G_{IJ,K}\dot{\p^I} Q^K \dot{Q^J} \right]  
\cr 
  &=& \int {\rm d}t \dn{3}{x} a^3\left[\PX \mathcal{D}_t Q^I G_{IJ}\mathcal{D}_t Q^J+\left(\mathcal{D}_I \mathcal{D}_J P + \PX {\mathcal R}_{IKLJ}\dot{\p^K}\dot{\p^L}
\right) Q^I Q^J\right]\, ,
  \label{example}
\end{eqnarray}
where we have discarded total derivatives. $\mathcal{D}_I$ denotes the covariant derivative associated to
$G_{IJ}$ (we thus have $\mathcal{D}_I \mathcal{D}_J P \equiv P_{,IJ}-\Gamma^K_{IJ} P_{,K}$) and ${\mathcal R}_{IJKL}$ is the Riemann tensor for $G_{IJ}$. 
 
The second-order action can be finally written in the rather simple form
\begin{eqnarray}
S_{(2)}&=& \half \int {\rm d}t\, \dn{3}{x} \,a^3 \left[ 
\left( \PX G_{IJ} + P_{,XX} \dot{\p_I} \dot \p_J \right) \mathcal{D}_t Q^I \mathcal{D}_t Q^J -\frac{\PX}{a^2}G_{IJ} \partial_i Q^I \partial^i Q^J 
\right.
 \cr
  && 
  \left.
  - M_{IJ}Q^I Q^J + 2 P_{,XJ} \dot \p_I Q^J \mathcal{D}_t Q^I  \right]\,,
\label{2d-order-action}
\end{eqnarray}
with the effective (squared) mass matrix
\begin{eqnarray}
 M_{IJ}  &=& -\mathcal{D}_I \mathcal{D}_J P - \PX \mathcal{R}_{IKLJ}\dot \p^K \dot
\p^L+\frac{X \PX}{H} (P_{,XJ}\dot \p_I+P_{,XI}\dot \p_J)
 \nonumber\\
 &&~~{} + \frac{X \PX^3}{2 H^2}(1-\frac{1}{c_s^2})\dot \p_I \dot \p_J-\frac{1}{a^3}\mathcal{D}_t\left[\frac{a^3}{2H}\PX^2\left(1+\frac{1}{c_s^2}\right)\dot \p_
I \dot \p_J\right]  \,.
\label{Interaction matrix}
\end{eqnarray}
The last term can be expanded, which yields
\begin{eqnarray}
M_{IJ}  &=& -\mathcal{D}_I \mathcal{D}_J P - \PX \mathcal{R}_{IKLJ}\dot \p^K \dot \p^L+\frac{X \PX}{H} \left(P_{,XJ}\dot \p_I+P_{,XI}\dot \p_J\right)
 \nonumber\\
  &&\hspace*{-4.0em}+ \dot \p_I \dot \p_J \left[-\frac{X \PX^3}{c_s^2 H^2}+  \frac{s \PX^2}{c_s^2}
  - \frac{3\PX^2}{2} \left(1+\frac{1}{c_s^2}\right)-\frac{\PX}{H}\left(1+\frac{1}{c_s^2}\right)\left(P_{,XX} \dot X+P_{,XK}\dot \p^K\right)\right] 
  \nonumber\\
  &&\hspace*{-4.0em}-\frac{\PX^2}{2H}\left(1+\frac{1}{c_s^2}\right)\left(\dot\p_I\mathcal{D}_t \dot \p_J +\dot\p_J \mathcal{D}_t \dot \p_I \right) \,.
\end{eqnarray}

The second-order action  (\ref{2d-order-action}) is one of the main results of the present work and generalizes more restrictive cases considered in the literature \cite{Deruelle:1991gy,Anderegg:1994xq,Seery:2005gb}. 
Let us now  discuss some of its features. First, one can notice that the 
non-flat nature of the field space metric $G_{IJ}$ manifests itself in the replacement of the 
ordinary time derivatives by covariant time derivatives $\mathcal{D}_t$ and of $P_{,IJ}$ by 
$\mathcal{D}_I \mathcal{D}_J P$, as well as the presence of the Riemann tensor $\mathcal{R}_{IJKL}$ in the matrix $M_{IJ}$ \cite{Sasaki:1995aw,bartjan}.

A second interesting consequence of (\ref{2d-order-action}) is how the non-trivial dependence of the 
initial action on the kinetic term $X$ affects the dynamics of the perturbations. One sees that 
the term involving the spatial gradients is simply rescaled by a factor $P_{,X}$ with respect to the standard case (where $P=X-V$ so that $P_{,X}=1$), which multiplies the metric $G_{IJ}$. The term quadratic in the time derivatives is changed in a subtler way, since instead of $P_{,X} G_{IJ}$, one finds
\beq
P_{,X} \tilde{G}_{IJ}\equiv P_{,X}\left(G_{IJ} + \frac{P_{,XX}}{\PX} \dot\p_I \dot \p_J\right)\,.
\eeq
This shows that the background velocity, of components $\dot{\p^I}$, represents a special direction in field space as far as the dynamics of the perturbations is concerned. This direction corresponds to the (instantaneous) adiabatic direction, which has been introduced in \cite{Gordon:2000hv} for multi-field inflation with standard kinetic terms. Introducing the `adiabatic' unit vector $e_1^I$, defined as 
\beq
\label{e1}
e_1^I=\frac{\dot{\p^I}}{\sqrt{2X}},
\eeq
and using (\ref{cs2}),
one finds
\beq
 \tilde{G}_{IJ}\equiv \frac{1}{c_s^2} e^1_I \, e^1_J 
 +\left(G_{IJ} -  e^1_I \, e^1_J\right),
\eeq
where the term between parentheses represents, in field space, the projection orthogonal to the adiabatic direction. 
This decomposition clearly shows that the adiabatic part of the perturbations, i.e. along  the field velocity, 
obeys a wave equation where the propagation speed is determined by the sound speed $c_s$, while the entropy perturbations, i.e. orthogonal to the scalar field velocity, propagate with the speed of light. This property, which was pointed out in \cite{Easson:2007dh,Huang:2007hh} for the specific case of two-field DBI inflation, turns out to be generic for any system governed by an action of the form (\ref{action}).

The decomposition into adiabatic and isocurvature modes can be made more explicit by introducing an orthonormal basis 
in field space, $\{e_n^I\}$  ($n=1, \dots, N$), where the first vector is the unit adiabatic vector introduced in 
(\ref{e1}).  The $N-1$ remaining vectors thus span the entropy subspace, which is orthogonal to the adiabatic direction. The decomposition of  the perturbations on this new basis reads 
\beq
Q^I= Q^n e_n^I,
\eeq
with an implicit summation on the index $n$. In order to replace the time derivatives of the $Q^I$ in terms 
of the time derivatives of the new components $Q^n$, one needs to take into account the time derivative of the basis vectors. It is in particular useful to define the quantities 
\beq
Z_{mn}= 
e_{mI} \mathcal{D}_t  e_{n}^I,
\label{Z}
\eeq
which satisfy the antisymmetry property $Z_{mn}=-Z_{nm}$ as a consequence of $\mathcal{D}_t(e^I_m e_{nI})=0$.
The second order action then reads
\begin{eqnarray}
 S_2 &=&   \half \int  {\rm d}t\, \dn{3}{x} \,a^3\  \left[ \frac{\PX}{c_s^2}(\dot{Q^1}+Z_{1m}Q^m)^2 +\PX \sum_{n>1}(\dot{Q^n}+Z_{nm}Q^m)^2 \right.
  \nonumber\\
  &-&\left. \frac{\PX}{a^2}\partial_iQ^n \partial^iQ^n -Q^n M_{nm} Q^m +2\sqrt{2X} (\dot{Q^1}+Z_{1m}Q^m)P_{,Xn}Q^n\right]
\label{2d-order-action-Qn}
\end{eqnarray}
where $M_{nm}\equiv e_n^{I} M_{IJ} e_m^{J}$ are the components of the interaction matrix in our new basis. Similarly, $P_{,Xn}\equiv e_n^I P_{,XI}$ and $P_{,n}\equiv e_n^I P_{,I}$.\\

So far, only  the first vector of the basis, $e_1$, has been specified. In the following, we will also specify
the second  element of the basis, $e_2^I$, as the unit vector pointing along the projection  on the entropy subspace of the field acceleration 
$\mathcal{D}_t \dot{\p^I}$. The  $\rm(N-2)$ remaining 
 vectors of the basis are left arbitrary in the present work. Note that \cite{bartjan} adopted a specific choice for all the vectors of the basis by considering the successive time derivatives of the background scalar fields.

\section{Curvature perturbations}
In the previous section, we have obtained the second order action in terms of the physical degrees of freedom 
$Q^I$. This completely determines the full dynamics of the linear perturbations. It is then useful to relate the quantities $Q^I$ to other perturbed quantities which can be of interest, in particular the gauge-invariant quantities describing the metric perturbations.

Instead of the metric written in the ADM form, we will work in this section with the (scalarly) perturbed 
FLRW metric written in the usual form (see \cite{ks,mfb} for detailed reviews on the theory of linear cosmological perturbations and e.g. \cite{Langlois:2004de} for a pedagogical introduction),
\beq
ds^2=-(1+2A)dt^2+2a\partial_i B dx^i dt+a^2\left[(1-2 \psi)\delta_{ij}+2\partial_{ij} E \right]dx^i dx^j\,.
\label{metric-generale}
\eeq
These metric perturbations can be combined to give the familiar gauge-invariant Bardeen potentials, defined by
 \begin{eqnarray}
 \label{Phi}
\Phi &\equiv& A - \frac{{\rm d}}
{{\rm d}t} \left[ a^2(\dot{E}-B/a)\right] \,,\\
 \label{Psi}
\Psi &\equiv& \psi + a^2 H (\dot{E}-B/a) \,.
\end{eqnarray}

For the matter, the  linear perturbations of the energy density, pressure and  momentum 
follow from the linearized energy-momentum, according to the expressions
\beq
\delta \rho \equiv -\delta T^0_0, 
\quad \delta P=\frac{1}{3}\delta T^i_i, 
\quad \partial_i \delta q=\delta T^0_i,
\eeq
while, in our case, the anisotropic stress vanishes. 
 The linear combination
 \begin{equation}
 \label{drhom}
\delta\rho_m = \delta\rho - 3H \delta q 
\end{equation} 
is gauge-invariant and defines the so-called comoving energy density perturbation (i.e. coincides with the energy density perturbation in the comoving gauge, characterized by $\delta q=0$).

Other useful gauge-invariant quantities are obtained by combining the metric and matter perturbations.
One can define the curvature perturbation on uniform-density
hypersurfaces
\begin{equation}
 \label{zeta}
- \zeta \equiv \psi + \frac{H}{\dot\rho}\delta\rho \,,
\end{equation}
and the so-called comoving curvature perturbation
\begin{equation}
 \label{R}
{\cal R} \equiv \psi - \frac{H}{\rho+P} \delta q \,.
\end{equation}
In the following, we will work in Fourier space for all these linear perturbations.
 
The gauge-invariant quantities defined above can be related to the physical degrees of freedom $Q^I$, which we have chosen to describe our system. The relations follow from 
the linearized Einstein's equations $\delta G_{\mu \nu}=\delta T_{\mu \nu}$, more precisely from  the constraints.
In particular,  the energy constraint reads
 \beq
3H(\dot \psi +H A)+\frac{k^2}{a^2}[\psi +H(a^2 \dot E -a B)]=-\half \delta \rho ,
 \label{energy-constraint}
 \eeq
while  the momentum constraint yields
\beq
\dot \psi +H A =-\half \delta q\, .
\label{momentum-constraint}
\eeq
The combination of these two constraints  gives a gauge-invariant relativistic generalisation of the Poisson equation, 
\beq
\frac{k^2}{a^2}\Psi=-\half \delta \rho_m\,.
\label{Poisson}
\eeq

Note that the $Q^I$'s correspond to the gauge-invariant  combinations $Q^I=\delta \p^I+ (\dot \p^I/ H)\, \psi$ where the $\delta \p^I$'s are the field perturbations in any gauge. In order to write the various quantities introduced above in terms of the $Q^I$, it is useful to notice that the ADM metric is equivalent to a metric of the form (\ref{metric-generale}) with the identification 
$A=\alpha$, $a B=\beta$, $\psi=0$ and $E=0$. Using (\ref{alpha}), one finds that 
the perturbed energy density and pressure, in the flat gauge, take the form
\beq
\delta \rho_{\rm flat} = \frac{\PX}{H c_s^2}\,\dot \p_I \,\mathcal{D}_t(H Q^I)+(2X P_{,XI}-P_{,I})Q^I\,,
\label{delta-rho}
\eeq
and
\beq
\delta P_{\rm flat}=\frac{\PX}{H}\, \dot \p_I \,\mathcal{D}_t (H Q^I)+P_{,I}Q^I.
\label{delta-P}
\eeq 
Moreover, substituting (\ref{alpha}) in (\ref{momentum-constraint}) yields the expression of $\delta q$, in the flat gauge, in terms of the $Q^I$. One thus gets
\beq
{\cal R} = \frac{H}{2X}\dot \p_I Q^I 
\label{R_QI}
\eeq
 and 
\beq
\delta\rho_m= \frac{\PX}{c_s^2}\dot{\p_I}\mathcal{D}_t Q^I +2X
P_{,XI} Q^I-P_{,I} Q^I-\frac{\PX}{H}\left(\frac{X \PX}{c_s^2}-3H^2\right)\dot{\p_I} Q^I .
\label{drhom_Q}
\eeq
Taking the time derivative of the expression (\ref{R_QI}) for  ${\cal R}$ and combining the result with  
(\ref{drhom_Q}) and (\ref{Poisson}), one obtains an important result for the time evolution of ${\cal R}$:
\beq
%\label{Rdot}
\dot{ \mathcal{R}}=\frac{H}{\dot H}\frac{c_s^2 k^2}{a^2}\Psi+\frac{H}{2 X\PX}
\left[(1+c_s^2)P_{,I}^{\perp} - 2X c_s^2 P_{,XI}^{\perp}\right] Q^I 
\eeq
where, for any covector $A_I$ in field space, $A_I^{\perp}\equiv A_I -(e_1^K A_K)e_{1I}$ represents its projection on the isocurvature subspace, i.e. orthogonal to the adiabatic direction.

All the above results can be rewritten in the basis $\{e^I_n\}$, which distinguishes the adiabatic and isocurvature perturbations. One can express the comoving curvature perturbation 
as
\beq
{\cal R} = \frac{H}{\sqrt{2X}}Q^1\, ,
\eeq
which illustrates that ${\cal R}$ characterizes the purely adiabatic part of the perturbations.
The comoving energy density perturbation now reads
\beq
  \delta\rho_m = \frac{\PX \sqrt{2X}}{c_s^2} \left[ \dot Q^1 +\left(\frac{\dot H}{H}-\frac{\dot X}{2X}\right)Q^1\right]+2 X\left(\sum_{n>1}P_{,Xn}Q^n\right)-\left(1+\frac{1}{c_s^2}\right) P_{,2}Q^2.
\eeq
Finally, the time evolution of ${\cal R}$ can be rewritten as 
\beq
\label{Rdot}
\dot{ \mathcal{R}}=\frac{H}{\dot H}\frac{c_s^2 k^2}{a^2}\Psi+\frac{H}{2 X\PX}\left[ (1+c_s^2)P_{,2}Q^2 - 2 X c_s^2\sum_{n>1}P_{,Xn}Q^n \right] \,.
\eeq
When one can ignore the first term on the right hand side, on sufficiently large scales, one recovers the familiar result that the curvature perturbation is sourced by entropy perturbations only. However, in our  case, this entropy source term contains two contributions. 

The first one, proportional to $P_{,2}$, is 
a generalization of the term obtained in \cite{Gordon:2000hv}. Indeed, for $P=X-V(\p^I)$ and  a flat field space metric, the above relation with two fields reduces to 
\beq
\dot{ \mathcal{R}}=\frac{H}{\dot H}\frac{k^2}{a^2}\Psi-\frac{H}{X} V_{,2} Q^2=\frac{H}{\dot H}\frac{k^2}{a^2}\Psi+\frac{2H}{\dot \sigma} \dot \theta \delta s\, ,
\quad \left(P=X-V,\ G_{IJ}=\delta_{IJ} \right)
\label{R-standard}
\eeq
where we have introduced, in the second equality,  $\dot \sigma \equiv \sqrt{2X}$, $\delta s=Q^2$ and $\dot \theta=-{V_{,2}}/{\dot \sigma}$, corresponding to the notation of \cite{Gordon:2000hv}.

 This result has been extended  in \cite{bartjan} to the multi-field situation, still for
 $P=X-V$ but with a general metric on field space.
In this case, the curvature perturbation on large scales is still sourced by a term related to the perturbation 
in the direction $e_2^I$, i.e. along the entropic projection of the field acceleration $\mathcal{D}_t \dot \p^I$. Therefore, on large scales, the curvature perturbation is not conserved  if  the trajectory in field space is not geodesic. By a geodesic 
trajectory, we mean here that  $\mathcal{D}_t \dot \p^I$ is  proportional to $\dot\p^I$.  Equivalently, in view of the background equations for the scalar fields (\ref{KG}), the curvature perturbation is not conserved on large scales if the gradient of $P$  along $e_2^I$ does not vanish.

In the more general situation considered here, one sees that the mixed derivatives $P_{,XI}$ along the entropy directions give an additional contribution to the non-conservation of the curvature perturbation. 
Therefore, if these mixed derivatives are nonzero, the curvature perturbation is not conserved on large scales, even if the motion in field space is geodesic.

When dealing with scalar fields, it is easier to work with the comoving curvature perturbation ${\cal R}$,  rather than with  the curvature perturbation on uniform density hypersurfaces $\zeta$.  
One can nonetheless write down the evolution equation for $\zeta$ in the form (see \cite{Wands:2000dp} and
\cite{Langlois:2005ii,Langlois:2005qp} for the exact and covariant form of this relation)
\begin{equation}
\label{eq:dotzeta}
\dot\zeta = -H \frac{\delta P_{\rm nad}}{\rho+P} - {\Sigma} \,,
\end{equation}
where $\delta P_{\rm nad}$ is the
non-adiabatic pressure perturbation, defined as
\beq
\delta P_{\rm nad}=\delta P-\frac{\dot P}{\dot \rho}\delta\rho\,,
\label{defPnad}
\eeq
and $\Sigma$ is
the scalar shear along comoving worldlines,
which is given explicitly by
\begin{eqnarray}
\hspace*{-0.5em} \frac{\Sigma}{H} &\equiv& - \frac{k^2}{3H} \left\{
\dot{E}-(B/a) + \frac{\delta
    q}{a^2(\rho+P)} \right\} \nonumber \\
\hspace*{-0.5em} &=& - \frac{k^2}{3a^2H^2} \zeta
 - \frac{k^2 \Psi}{3a^2H^2} \left[ 1 + \frac{2\rho}{9(\rho+P)}
     \frac{k^2}{a^2 H^2} \right]. \nonumber \\
\end{eqnarray}
In our case, the non-adiabatic pressure perturbation can be expressed as 
\beq
\delta P_{nad}= \frac{\delta \rho_m}{6HX \PX }\left[(1+c_s^2)P_{,I} - 2X c_s^2 P_{,XI}\right]\dot \p^I +\left[(1+c_s^2)P_{,I}^{\perp} - 2X c_s^2 P_{,XI}^{\perp}\right]Q^I.
\label{Pnad-exact}
\eeq
On sufficiently large scales, one can neglect the first term, proportional to the comoving energy density, and only the second term on the right hand side contributes to the non-adiabatic pressure perturbation.

\section{Two-field case}

We now specialize our formalism to the case where only two scalar fields are present. In this context, the 
entropy subspace is one-dimensional and the basis $\{ e_1^I, e_2^I\}$ is completely specified. In order to make a direct comparison with the previous literature, we will use a more traditional notation, and replace the
subscripts $1$ and $2$ by respectively $\sigma$ and $s$, so that
\beq
 Q^1\equiv Q_\s, \qquad Q^2\equiv Q_s. 
\eeq 

\subsection{Background equations}

The background equations of motion for the scalar fields can  be decomposed into  adiabatic and 
entropic  equations. Defining  
\beq\dot \sigma\equiv \sqrt{2X}\,,
\eeq
the adiabatic equation of motion can be written as 
\beq
\ddot \s+\left(3H+\frac{\dot \PX}{\PX}\right)\dot \s - \frac{1}{\PX} P_{,\s}=0\,, \qquad P_{,\s}\equiv e^I_\s P_{,I}.
\eeq
Moreover, by using the decomposition
\beq
\dot \PX=P_{,XX}\dot X+P_{,XI}\dot\phi^I=P_{,XX}\,\dot\s\,\ddot\s+P_{,X\s}\,\dot\s,
\eeq
the adiabatic equation of motion can also be rewritten as 
\beq
\ddot\sigma=c_s^2\left(\frac{P_{,\s}}{P_{,X}}-\frac{\dot\s^2 P_{,X \s}}{\PX}-3H\dot\s\right).
\eeq
The entropy part of the equations of motion gives the rate of change of the adiabatic basis vector 
$e_\s^I$ in terms of the entropy basis vector $e_s^I$:
\beq
{\mathcal D}_t e^I_\s=\frac{P_{,s}}{\PX \dot\sigma}e^I_s.
\eeq
This  is the generalization of the equation giving the time derivative of  the angle $\theta$ between the initial field basis and the adiabatic/entropy basis. This also implies that the non vanishing components of the matrix 
$Z_{mn}$ are given here by
\beq
Z_{s \s}=-Z_{\s s}=\frac{P_{,s}}{\PX \dot\s}.
\eeq

To manipulate the equations of motion for the perturbations, the following identities will also be useful:
\beq
\dot P_{,\sigma}=\frac{P_{,s}^2}{\dot\s \PX}+P_{,\s\s}\dot\s+P_{,X\s}\dot\sigma\ddot\sigma,
\eeq

\beq
\dot P_{,s}=-\frac{P_{,s} P_{,\sigma}}{\dot\s \PX}+P_{,\s s}\dot\s+P_{,Xs}\dot\sigma\ddot\sigma. 
\eeq

\subsection{Equations of motion for the perturbations}
Specializing our action (\ref{2d-order-action-Qn}) to the present two-field case, one can express the 
action in terms of the quantities $Q_\s$ and $Q_s$ and easily derive their equations of motion.
The adiabatic equation of motion can be written in the compact form
\begin{eqnarray}
\label{Qsigma}
 \ddot{Q}_{\sigma}&+&\left[3H+\frac{c_s^2}{\PX}\Tdot{\left(\frac{\PX}{c_s^2}\right)}\right]
 \dot{Q}_{\sigma}+\left(\frac{c_s^2 k^2}{a^2}+\mu_{\s}^2\right)  Q_{\sigma} \, 
\cr
&=&
\Tdot{\left(\Xi Q_s\right)}
-\left(\frac{\Tdot{(H c_s^2)}}{H c_s^2}
-\frac{P_{,\sigma}}{\dot \s \PX}\right)  \Xi\, Q_s\,, 
\end{eqnarray}
with
\beq
\Xi\equiv \frac{1}{\dot \s\PX}\left[(1+c_s^2)P_{,s} -c_s^2 P_{,Xs}\dot\s^2\right]
\eeq
and
\begin{eqnarray}
\mu_{\s}^2 &\equiv & -\frac{\Tddot{(\dot\s/H)}}{\dot\s/H}-\left(3H+\frac{c_s^2}{\PX}\Tdot{\left(\frac{\PX}{c_s^2}\right)}+\frac{\Tdot{(\dot\s/H)}}{\dot\s/H}\right)
\frac{\Tdot{(\dot\s/H)}}{\dot\s/H}
\label{mu_sigma}\, .
\end{eqnarray}
The equation of motion for the entropy part is given by 
\begin{eqnarray}
 \ddot{Q}_s+\left(3H+\frac{\dot \PX}{\PX}\right)\dot{Q}_s+\left(\frac{ k^2}{a^2}+\mu_s^2\right)Q_s
  &=&-\frac{\Xi}{c_s^2}\left[\dot Q_{\sigma} -H\left(\frac{\PX \dot \s^2}{2 H^2}+\frac{\ddot \s}{H \dot \s}\right)Q_{\sigma}\right]
  \nonumber\\
  &=&-\frac{\dot\sigma}{H c_s^2}\, \Xi\,  \dot{\mathcal{R}}\, .
\label{delta_s_all_scales}
\end{eqnarray} 
with
\beq
\mu_s^2\equiv -\frac{P_{,ss}}{\PX}+\half \dot \s^2\tilde{R}-\frac{1}{2 c_s^2 X}\frac{P_{,s}^2}{\PX^2}+2\frac{P_{,Xs} P_{,s}}{\PX^2}\,,
\eeq
where
$\tilde{R}$ denotes the Riemann scalar curvature of the field space. 

Using the relation (\ref{Rdot}), this equation of motion can also be rewritten as 
\beq
\ddot{Q}_s+\left(3H+\frac{\dot \PX}{\PX}\right)\dot{Q}_s+\left(\frac{ k^2}{a^2}+\mu_s^2+\frac{\Xi^2}{c_s^2}\right) Q_s=-\frac{\dot\s}{\dot H}\Xi \frac{k^2}{a^2} \Psi\,,
\label{delta_s}
\eeq
which can be useful on large scales, when the right hand side can be neglected. In this limit, the above equation shows that the entropy perturbation $Q_s$ evolves independently of the adiabatic mode.

It is clear from the previous equations that $\Xi$ quantifies the coupling between the entropy  and  adiabatic modes. One also notices that the equations of motion are characterized by a friction term that differs from the usual Hubble friction term. The additional friction term for the entropy perturbation is the same as the one which appears in the background equation of motion, but different from the friction term for the adiabatic perturbation. Finally, as already emphasized for $N$ fields, we observe that the adiabatic mode propagates with the speed of sound $c_s$, while the isocurvature mode 
is  characterized by a propagation with the usual speed of light.

For a  Lagrangian of the form
\beq
\label{P_standard}
P=X-V(\p^I), 
\eeq
the mixing parameter reduces to $\Xi= -2 V_{,s}/\dot \s=-2Z_{\sigma s}$. This shows that, in this particular case, the coupling between the adiabatic and entropy modes is directly related to the `rotation'  of the 
adiabatic/entropy basis in field space, i.e. to the bending of the background trajectory in field space. It is worth noting that this direct link no longer holds in the general case because of the term 
proportional to $P_{,Xs}$ in $\Xi$.
When the field metric is flat, $G_{IJ}=\delta_{IJ}$, one can introduce, as in \cite{Gordon:2000hv}, the rotation angle bewteen the initial basis and the adiabatic/entropy basis. One thus finds that $\Xi=2\dot\theta$. For a Lagrangian of the form (\ref{P_standard}) but with the field metric
\beq
G_{IJ}\, d\phi^I\, d\phi^J=d\phi^2+e^{2b(\phi)} d\chi^2,
\label{metric_polar}
\eeq
one finds  \cite{DiMarco:2002eb}  that $\Xi=2\dot\theta+b'\dot\s \sin\theta$, where 
we see now that the additional term simply comes from the non-trivial covariant derivative in (\ref{Z}). 
It is also easy to check that our  equations (\ref{Qsigma}-\ref{delta_s})  reduce to the results of \cite{Gordon:2000hv,DiMarco:2002eb} in these particular cases. Note that, in the first case, non-linear extensions of these equations have been obtained in \cite{Langlois:2006vv}, based on a covariant formalism similar to that of \cite{Langlois:2005ii,Langlois:2005qp}.

When one can neglect spatial gradients, the expression (\ref{Rdot}) reduces to
\beq
\dot{Q_{\sigma}}+\left(\frac{\dot{H}}{H}-\frac{\ddot{\sigma}}{\dot{\sigma}}\right) Q_{\sigma}-  \Xi \, Q_s \approx 0\,,
\label{Q1-large-scales}
\eeq
which implies that there exists a first integral for $Q_{\sigma}$ and that the second-order equation of motion (\ref{Qsigma}) is not necessary in this limit. Indeed one can check that the large-scale limit of (\ref{Qsigma}) is a consequence of (\ref{Q1-large-scales}).

To conclude this section, let us give the approximate form of the evolution equations for the perturbations when the spatial gradients can be neglected:
\beq
\dot{ \mathcal{R}} \approx \frac{H}{\dot \s}\, \Xi \, Q_s \,,
\label{R-large-scales}
\eeq
\beq
\ddot{Q_s}+\left(3H+\frac{\dot \PX}{\PX}\right)\dot{Q_s}+\left(\mu_s^2+\frac{\Xi^2}{c_s^2}\right) Q_s \approx 0\,.
\label{Q2-large-scales}
\eeq

\section{Quantum fluctuations}
We now consider the quantization of the system discussed in the previous section. In order to do so, we follow the usual procedure for single field inflation. The first step  consists in introducing a new variable which is canonically normalized, with conformal time, and whose effective mass is time
dependent because of the expansion of the Universe. 
In the present case, we have two degrees of freedom. By introducing the new  fields
\beq
v_{\s}=\frac{a \sqrt{\PX}}{c_s}\, Q_{\s} \,,\qquad \,v_{s}=a\,\sqrt{\PX}\, Q_s\,,
\label{v}
\eeq
respectively for the adiabatic and entropy degrees of freedom, one can rewrite the second-order action in the form
\begin{eqnarray}
\label{S_v}
S_{(2)}=\frac{1}{2}\int {\rm d}\tau {\rm d}^3k &&\left[ 
  v_\s^{\prime\, 2}+ v_s^{\prime\, 2} -2\xi v_\s^\prime v_s-k^2 c_s^2 v_\s^2 -k^2 v_s^2 
\right. \cr
&& \left.
+\Omega_{\s\s}v_\s^2+\Omega_{ss} v_s^2+2\Omega_{s\s}v_\s v_s\right]
\end{eqnarray}
with
\beq
\xi=\frac{a}{c_s}\Xi=\frac{a}{\dot \s \PX c_s}[(1+c_s^2)P_{,s}-c_s^2 \dot \s^2 P_{,Xs}]\,,
\label{11}
\eeq
and
\beq
\Omega_{\s\s}=\frac{z''}{z}\,,\qquad  \Omega_{s\s}=\frac{z'}{z}\xi\,, \qquad
\Omega_{ss}=\frac{\alpha''}{\alpha}-a^2 \mu_s^2\,,
\eeq
where we have introduced the two background-dependent  functions 
\beq
z=\frac{a \dot \s \sqrt{\PX}}{c_s H}, \qquad \alpha=a\sqrt{\PX}\,,
\eeq
and used the conformal time $\tau$ defined by $\tau = \int {dt}/{a(t)}$.

For single field inflation with a standard kinetic term, $z$ reduces to the usual function $a\dot\s/H$.
In the general case considered here, the function $z$ is still the ratio between $v_\s$ and the comoving curvature perturbation defined in the previous section, i.e.
\beq
v_{\s}=z \, \mathcal{R}.
\eeq

The equations of motion derived from the action (\ref{S_v}) can  be written in the compact form
\begin{eqnarray}
v_{\s}''-\xi v_{s}'+\left(c_s^2 k^2-\frac{z''}{z}\right) v_{\s} -\frac{(z \xi)'}{z}v_{s}&=&0\,.
\label{eq_v_sigma}
\\
v_{s}''+\xi  v_{\s}'+\left(k^2- \frac{\alpha''}{\alpha}+a^2\mu_s^2\right) v_{s} - \frac{z'}{z} \xi v_{\s}&=&0\,.
\end{eqnarray}
As mentioned already, the above system shows clearly that the adiabatic degree of freedom $v_\s$ is sensitive to the sound horizon, while the entropy degree of freedom $v_s$ is sensitive to the usual Hubble radius. 
In the general case where $\xi$ is non-vanishing, the above system is coupled and we leave for a future investigation the analysis of the perturbations generated by this coupled system. In the following, we
will just consider the simple case where $\xi$ can be neglected while the perturbations cross the Hubble radius and sound horizon. This is similar to the situation considered in \cite{Huang:2007hh} for two-field DBI inflation.

When $\xi=0$, the system is completely decoupled and one can analyse separately the adiabatic and entropy 
degrees of freedom.  For completeness, let us repeat  the standard analysis (see \cite{mfb}, or \cite{Langlois:2004de}). Each variable $v$, either $v_{\s}$ or $v_s$,  becomes 
a quantum field which is decomposed as 
\beq
\label{Fourier_quantum}
\hat v (\tau, \vec x)={1\over (2\pi)^{3/2}}\int {\rm d}^3k \left\{{\hat a}_{\vec k} v_{k}(\tau) e^{i \vec k.\vec x}
+ {\hat a}_{\vec k}^\dagger v_{k}^*(\tau) e^{-i \vec k.\vec x} \right\},
\eeq
where  the $\hat a^\dagger$ and  $\hat a$ are 
 creation and annihilation operators (there are two sets of creation and annihilitation operators, one for the adiabatic degree of freedom, the other for the entropy one), which satisfy the 
usual commutation rules 
\beq
\label{a}
\left[ {\hat a}_{\vec k}, {\hat a^\dagger}_{\vec k'}\right]= \delta(\vec k-\vec k')\, ,
\quad
\left[ {\hat a}_{\vec k}, {\hat a}_{\vec k'}\right]= 
\left[ {\hat a^\dagger}_{\vec k}, {\hat a^\dagger}_{\vec k'}\right]= 0\, .
\eeq
The action with $\xi=0$ implies that  the conjugate momenta for $v_{\s}$ and $v_s$ are respectively 
$v'_{\s}$ and $v'_s$. Therefore, in either case, the canonical quantization  
for $\hat v$ and its conjugate momentum leads to the condition
\beq
v_{k} {v'_{k}}^*-v_{k}^*v'_{k}=i\,.
\label{wronskien}
\eeq
which must be satisifed by the complex function $v_k(\tau)$.

For the adiabatic degree of freedom, the function  $v_{\s\, k}(\tau)$ satisfies the equation of motion (\ref{eq_v_sigma}) with $\xi=0$, i.e.
\beq
v_{\s}''+\left(c_s^2 k^2-\frac{z''}{z}\right) v_{\s} =0.
\eeq
In the slow-roll limit, and when $c_s$ varies sufficiently slowly while the scale of interest crosses out the sound horizon, one can 
take the approximation $z''/z\simeq 2/\tau^2$, so that the general solution is known analytically within this approximation. Finally, we require that the solution on small scales behaves like the Minkowski vacuum. This leads to the solution 
\beq
v_{\s\, k}\simeq  \frac{1}{\sqrt{2k c_s}}e^{-ik c_s \tau }\left(1-{i\over k c_s\tau}\right),
\eeq
where the normalization is imposed by the condition (\ref{wronskien}).
This implies that the power spectrum of the adiabatic fluctuations is given by
\beq
\label{power_sigma}
{\cal P}_{Q_\s}=\frac{k^3}{2\pi^2}|v_{\s\, k}|^2\frac{c_s^2}{a^2 P_{,X}}\simeq\frac{H^2}{4\pi^2 c_s P_X},
\eeq
where the quantities on the right hand side are evaluated at 
 the {\it sound horizon crossing}. This can be translated into the power spectrum of the curvature perturbation 
 ${\cal R}$, 
\beq
{\cal P}_{\cal R}=\frac{k^3}{2\pi^2}\frac{|v_{\s\, k}|^2}{z^2}\simeq\frac{H^4}{8\pi^2 c_s X P_X}=\frac{H^2}{8\pi^2 \epsilon c_s}\,,
\eeq
where $\epsilon \equiv -\dot H/H^2$. In the decoupled case, the adiabatic sector is equivalent to the single field \textit{k}-inflation scenario and 
the above expression coincides with the result given in \cite{Garriga:1999vw}. 
A more refined treatment, taking into account the next-to-leading-order corrections, can be found in 
the appendix of \cite{Chen:2006nt}.

Let us now consider the entropy degree of freedom. When $\xi=0$, the complex function  $v_{s\, k}(\tau)$ satisfies the equation of motion 
\beq
v_{s}''+\left(k^2- \frac{\alpha''}{\alpha}+a^2\mu_s^2\right) v_{s}=0
\eeq
In the slow-roll limit,  $\epsilon \equiv -\dot H/H^2$ is a small coefficient and its time derivative 
is at least second order in slow-roll. 
Since 
\beq
\dot \epsilon=2 H \epsilon\left(\epsilon+\frac{\Tdot{ (X \PX)}}{2H X \PX}\right),
\eeq 
one must consider  ${\dot \PX}/(H \PX)$ as first order in slow-roll.
 One can then write, around the time of Hubble-crossing ($k=a H$),
\beq
v_s''+k^2 v_s-\frac{1}{\tau^2}\left(\nu_s^2-\frac{1}{4}\right)v_s=0
\label{eq_v_s}
\eeq
with
\beq
\nu_s^2\simeq\frac{9}{4}-\frac{3\dot H}{H^2}+\frac{3\dot \PX}{
2H \PX}+\frac{P_{,ss}}{\PX H^2}-\frac{\dot \s^2}
{2 H^2}\tilde{R}-(2+\frac{1}{c_s^2})\frac{P_{,s}^2}{ \dot \s^2H^2 \PX^2}\,,
\eeq
where we have neglected the square and the time derivative of $\dot\PX/(H \PX)$. 
 
The solution of (\ref{eq_v_s}) with the appropriate asymptotic behaviour is 
\beq
v_s=\frac{\sqrt{\pi}}{2}e^{i(\nu_s+1/2)\pi/2}\sqrt{-\tau}\, H_{\nu_s}^{(1)}(-k\tau),
\eeq
where $H_{\nu}^{(1)}$ is the Hankel function of the first kind of order $\nu$. 
Note that we did not assume 
 that the last three terms were small. If they are big,  the effective mass is important and the entropy fluctuations are suppressed.  The last term in particular shows that the effect of the bending of the background trajectory on the entropy mode  is enhanced in the small sound speed limit.

If the last three terms are small, i.e. in the limit 
$\nu_s \rightarrow 3/2$, one finds 
\beq
v_{s\, k}\approx  \frac{1}{\sqrt{2k}}e^{-ik \tau }\left(1-\frac{i}{ k \tau}\right).
\eeq
In this case, the entropy power spectrum, at Hubble crossing,  is given by 
\beq
\label{power_s}
{\cal P}_{Q_s}=\frac{k^3}{2\pi^2}\frac{|v_{s\, k}|^2}{a^2 P_{,X}}\simeq \frac{H^2}{4\pi^2 P_{,X}},
\eeq
where the quantities are evaluated at {\it Hubble crossing}. One sees 
that the ratio between the entropy and adiabatic power spectra, evaluated at their respective `horizon'  crossings,  is therefore the speed of sound $c_s$. 

\section{Conclusion}
\label{sec:conclusion}
In this work, we have studied the linear perturbations for a very large class of multi-field inflationary models, which includes most of the multi-field models studied so far. Our formalism should  also be useful for studying the perturbations in multi-field inflationary models that will be constructed in the future. In order to obtain the equations of motion for the perturbations, as well as their quantum initial conditions, we have calculated directly the action at second order in the linear perturbations. To facilitate the analysis of the equations of motion, we have decomposed the linear perturbations into one adiabatic mode, along the background inflationary trajectory, and $N-1$ entropy modes. We have shown that a generic feature of these models is that the adiabatic and entropy modes obey coupled wave equations, which propagate with the speed of light for all entropy modes but with the speed of sound $c_s$ for the adiabatic mode.   

We have also used this decomposition into adiabatic and entropy modes, to show that the (comoving) curvature perturbation, proportional to the adiabatic perturbation, is sourced by a very specific combination of the entropy perturbations. In this  combination, one finds a term which is non-zero when the  background trajectory in field space is bent, or, more precisely, when it is non-geodesic (with respect to the field space metric $G_{IJ}$) but we have also found another term, proportional to $P_{,Xs}$, which is not present for models with canonical kinetic terms. 

Interestingly, despite (or maybe thanks to) the fact that we have considered a much larger class of models than previous works, we have been able, in the two-field case, to write down the equations of motion in a very compact and simple form. 

Finally, it would be interesting to extend this work in several directions. The analysis of the quantum fluctuations when the adiabatic and entropy modes are coupled would be particularly interesting. It is more complicated than in the standard case, because the quantum-classical transition occurs in principle at different times for adiabatic and entropy modes with the same wave number $k$: at the usual Hubble crossing ($k=aH$) for the isocurvature modes, but at the crossing of the sound horizon ($kc_s=aH$) for the adiabatic mode.

In the present work, we have considered only linear perturbations. It would be interesting to go beyond the linear order and to study the non-Gaussianities that could be generated in this class of models. One can expect that the non-Gaussianities in this class of models would combine the types of non-Gaussianities exhibited in single-field DBI inflation and in multi-field inflation.

\vspace{0.3cm}
{\bf Acknowledgment:} We would like to thank V. Mukhanov, D. Steer, G. Tasinato, and K. Turzynski for  very instructive discussions.  

\section*{Appendix: multi-field DBI inflation}

An important example of inflation with non canonical kinetic term is the so-called DBI inflation model, where the  inflaton field corresponds to the position of a probe D3-brane moving in a warped background and its dynamics is described by a DBI action. Very recently, a few papers have considered a multifield extension of DBI inflation. Taking into account the angular position of the brane in addition to its radial position, they started from an effective Lagrangian of the form
\beq
P(X,\p^I)=-\frac{1}{f} \left(\sqrt{1-2 f X}-1\right)-V, \qquad X=-\frac{1}{2}G_{IJ}\nabla_{\mu} \p^I  \nabla^{\mu} \p^J
\label{P_DBI}
\eeq
where $f$ and $V$ are functions of the scalar fields. The above Lagrangian belongs to the class of models studied in the present work, and the results given in the main text can thus be applied to this particular case and compared with the results of \cite{Easson:2007dh,Huang:2007hh}.

The  speed of sound derived from  (\ref{P_DBI}) is given by 
\beq
c_s=\sqrt{1-2 f X}\equiv \frac{1}{\gamma},
\label{cs}
\eeq
where $\gamma$ is the analogue of the relativistic Lorentz factor. By taking the derivative of (\ref{P_DBI}) with respect to the kinetic term $X$, 
one sees that the DBI action is a very special case, which satisfies the identity
\beq
P_{,X}=\frac{1}{c_s}=\gamma.
\eeq
Since both $P_{,X}$ and $c_s$ appear very often in our equations for the perturbations, this special relation will bring simplifications. For example, the background coefficients multiplying $\dot{Q}_\s$ and $\dot{Q}_s$ become, respectively, $(3H+3\dot\gamma/\gamma)$ and $(3H+\dot\gamma/\gamma)$, in agreement with the equations given in   \cite{Easson:2007dh}. Similarly, the coefficients of $\dot{Q}_\s^2$ and $\dot{Q}_s^2$ in the action, 
which are respectively $a^3 P_{,X}/(2 c_s^2)$ and $a^3 P_{,X}/2$ reduce to 
$a^3/(2 c_s^3)$ and $a^3/(2 c_s)$ in the DBI case, as obtained in \cite{Huang:2007hh}.

We have also checked that, when specialized to the Lagrangian (\ref{P_DBI}) and the field metric 
(\ref{metric_polar}), our equations for the perturbations agree with the equations given in 
\cite{Easson:2007dh}. Note, however, that the background coefficients in the equations of \cite{Easson:2007dh} are expressed in terms of the angle $\theta$ between the initial basis and the adiabatic/entropy basis\footnote{This angle is denoted $\alpha$ in \cite{Easson:2007dh}.} in a form which makes them look singular when $\theta=0$. In our case, we obtain directly the equations of motion with well-behaved coefficients.

As far as the primordial power spectra are concerned, it is easy to verify that our expressions (\ref{power_sigma}) and 
(\ref{power_s}) reduce to the results of \cite{Huang:2007hh}, when using (\ref{cs}).

Let us finally note that the non standard contribution to the mixing coefficient, proportional to $P_{,Xs}$ is non zero in general in this model since
\beq
P_{,Xs}=\frac{X f_{,s}}{(1-2f X)^{3/2}}.
\eeq
This term vanishes only when $f_{,s}$ vanishes. In the usual case where $f$ is a function of the scalar field corresponding to the radial direction, $f_{,s}$ vanishes only if the background trajectory is strictly radial.

\end{document}